\documentclass[12pt,preprint]{aastex}
\begin{document}

\title{An Arc of Young Stars in the Halo of M82\altaffilmark{1}}

\author{T. J. Davidge}

\affil{Herzberg Institute of Astrophysics,
\\National Research Council of Canada, 5071 West Saanich Road,
\\Victoria, B.C. Canada V9E 2E7\\ {\it email: tim.davidge@nrc.ca}}

\altaffiltext{1}{Based on observations obtained with the
MegaPrime/MegaCam, a joint project of the CFHT and CEA/DAPNIA,
at the Canada-France-Hawaii Telescope (CFHT), which is operated by
the National Research Council (NRC) of Canada, the Institut National des
Sciences de l'Univers of the Centre National de la Recherche
Scientifique (CNRS) of France, and the University of Hawaii.}

\begin{abstract}

	The properties of the brightest resolved stars in an arc that was originally 
identified by Sun et al. (2005) and is located in the extraplanar regions of M82 
are discussed. The stars form an elongated structure that is traced over a projected area 
of $3.0 \times 0.8$ kpc. The integrated brightness is M$_V \sim -11$, while the 
total stellar mass is between $3 \times 10^5$ M$_{\odot}$ and $2 \times 10^6$ M$_{\odot}$. 
If there is only foreground extinction then the youngest stars 
have a metallicity $Z \geq 0.008$ and an age log(t$_{yr}) \sim 7.75$; 
thus, the youngest stars formed at roughly the same time as stars in tidal features that 
are associated with other M81 Group galaxies. If the arc is dispersing 
then it will deposit young, chemically enriched stars into the M82 halo.

\end{abstract}


\section{INTRODUCTION}

	As one of the closest galaxy groups, the M81 Group is an important 
laboratory for probing galaxy evolution. Unlike the Local Group, the M81 Group contains 
galaxies that have experienced cosmologically recent interactions, with the elevated levels 
of star formation in M82 and NGC 3077 likely triggered by an encounter with M81 within the 
past few hundred million years (Brouillet et al. 1991; Yun, Ho, \& 
Lo 1994). Subsequent studies of the stellar fossil record in M82 
have since found evidence of wide-spread elevated levels of star-forming 
activity roughly 0.5 - 1 Gyr in the past (e.g. de Grijs, O'Connell, \& Gallagher 2001; 
Mayya et al. 2006). More recent epsiodes of star-forming activity have been restricted to 
the inner regions of M82 (e.g. Gallagher \& Smith 1999; Forster Schreiber 
et al. 2003; Smith et al. 2006). 

	The interactions in the M81 Group have had a major 
impact on the intracluster environment. Feedback from supernovae power an outflow 
from M82 that injects chemically enriched material into the intracluster medium 
(e.g. Shopbell \& Bland-Hawthorn 1998), which in turn may 
interact with surrounding clouds (e.g. Devine \& Bally 1999). 
Tidal interactions can also pull material from galaxies, and 
morphogical signatures of this activity are seen near some 
M81 Group galaxies (e.g. Karachentseva, Karachentsev, \& Boerngen 1985). 
The molecular material in M82 has been severely disrupted (Walter, Weiss, \& Scoville 
2002), and the morphology of the galaxy has been affected; while it now has an amorphous 
appearance, there are indications that it may have been a late-type spiral or irregular 
galaxy before the encounter with M81 (e.g. O'Connell \& Mangano 1978).

	Sun et al. (2005) find an arc-like feature in the southern regions of 
M82, which they refer to as `M82 South'. This object has a flat spectral energy 
distribution (SED) at visible wavelengths, a relatively high surface brightness, and 
is also seen in the UV (Figure 2 of Hoopes et al. 2005). 
M82 South is located just outside of the region 
imaged for the M82 Surveys Mosaic with the HST ACS (Mutchler et al. 2007). 
It can be anticipated that at least some of the light from M82 South originates 
from young stars, and resolving stars in such a feature would be of great interest. 
The light that has been detected from other sources in the extraplanar regions of M82 
appears to be from excited gas, and not stars. The detection of stars in M82 South would 
thus make it the first stellar structure to be detected in the outer regions of M82. 
Moreover, studies of the spatial distribution of stars in M82 South 
will also provide information for probing its structural properties and origins. 
Finally, if M82 South contains young stars and is not 
a gravitationally bound structure then it may contribute to the stellar content of the M82 
halo. During the past decade it has been demonstrated that the outer regions of nearby 
spiral galaxies contain stars spanning a range of ages and metallicities (e.g. Brown et al. 
2006; Mouchine 2006), and the disruption of structures like M82 South will 
diversify the stellar content of the halos of interacting galaxies. 
In this letter, we report on the detection of young stars in M82 South. It is shown that the 
youngest stars in M82 South have ages that are comparable to those of stars in 
tidal features near other M81 Group galaxies.

\section{OBSERVATIONS}

	The data were obtained with the MegaCam imager (Boulade et al. 2003) on the 
3.6 metre Canada-France-Hawaii Telescope (CFHT) as part of a wide-field survey of M81 and 
M82. The detector in MegaCam is a mosaic of 36 $2048 \times 4612$ pixel$^2$ CCDs, and 
each exposure covers roughly 1 degree$^2$ with 0.185 arcsec pixel$^{-1}$. 
Four 360 second exposures were recorded through $r'$ and $i'$ filters 
with the midpoint between M82 and M81 centered on the detector mosaic. 
Stars in the final images have 0.8 arcsec FWHM in $r'$ and 0.7 arcsec FWHM in $i'$.

	The raw data were processed with the ELIXER package at the CFHT, and this included 
bias subtraction, flat-fielding, and fringe removal.
The reduced images were then aligned and combined by the author at the HIA. 
The photometric measurements were made with ALLSTAR (Stetson \& Harris 1988), using a 
point-spread function (PSF) constructed with the DAOPHOT (Stetson 1987) PSF routine. 
The photometric calibration is based on the zeropoints 
that are placed in MegaCam data headers as part of the ELIXER processing. These 
zeropoints are calculated from standard star observations that are recorded 
as part of the MegaCam queue observing process.
 
\section{RESULTS}

\subsection{The Morphology and Integrated Brightness of M82 South}

	M82 South was discovered independently by the author when the locations of 
stars that were photometered with DAOPHOT were plotted (Davidge 2008 in preparation). 
A section of the final $i'$ MegaCam image that includes M82 South is shown in Figure 1. 
An expanded view of the region near M82 South is shown in the lower right hand 
corner of Figure 1, while the spatial distribution of sources in this same area with 
$i'$ between 22.5 and 24.5, which is the magnitude range where stars associated with M82 
South are seen (\S 3.2), is shown in the lower left hand corner. M82 South 
is 5.4 arcmin from the center of M82 ($\sim 6$ kpc projected 
distance if $\mu_0 = 27.9$ -- Sakai \& Madore 1999), 
and its stars form a band with a width of $\sim 40$ arcsec ($\sim 0.8$ kpc). 
A prominent strip of diffuse light falls near the southern edge of the 
stellar distribution. The stellar density in the richest part of M82 South is 
$\sim 120$ arcmin$^{-2}$, whereas in the areas marked `Background' in Figure 1 
the mean stellar density is at least $\sim 6 \times$ lower. The Western end of M82 South 
is marked by a distinct drop in stellar density, while there may be a tendency for the 
stars in M82 South to broaden or curve north at the eastern edge of Figure 1.

	The integrated brightness within $\pm 20$ arcsec of the M82 South ridgeline 
is $r' \sim 16.6 \pm 0.5$, so that M$_V \sim -11 \pm 0.5$, and the mean surface brightness 
is $\sim 26$ mag arcsec$^{-2}$. To make this measurement, 
the background was measured on both sides of M82 South, in the regions marked in 
Figure 1, and the results were averaged to obtain a background level that is 
appropriate for the ridgeline of M82 South. If M82 South is a simple stellar system 
with an age log(t) $\sim 7.75$ (see below), then log(M/L$_V) \sim -0.85$, 
(e.g. Mouchine \& Lancon 2003), where M/L$_V$ is the mass-to-light ratio in $V$; 
the total mass is then $(3 \pm 2) \times 10^5$ M$_{\odot}$. This is a lower limit, since 
M/L$_V$ will be higher if an older population is present. To estimate an upper 
mass limit M/L$_V = 1$ was assumed, and the total mass in this case is 
$(2 \pm 1) \times 10^6$ M$_{\odot}$.

\subsection{Stars in M82 South}

	The $(i', r'-i')$ CMDs of objects within $\pm 20$ 
arcsec of the M82 South ridgeline are shown in the middle row of Figure 2. 
The CMDs of the background/control fields indicated in Figure 1 
are also shown. The control fields to the north east and south west of M82 South 
contain very different numbers of stars, due to the gradient in stellar density in the 
outer regions of M82, and this gradient makes it difficult to place the northern boundary 
of M82 South with confidence. With the caveat that the northern control field may contain 
some stars belonging to M82 South, then it appears that no more than one third of the 
sources within $\pm 20$ arcsec of the M82 South ridgeline do not belong to M82 South.

	Contamination from foreground Galactic stars, background 
galaxies, and stars in the M82 field can be accounted for 
statistically by assessing the color and brightness distributions of objects in M82 South. 
The net $r'-i'$ color function of sources with $i'$ between 
23.5 and 24.5 (M$_{i'}$ roughly between --4.5 and --3.5), constructed by subtracting 
the mean color function of the control fields from that in M82 South, and the net $i'$ 
luminosity function (LF) of objects in M82 South, constructed by subtracting the mean LF of 
sources in the control fields from that in M82 South, are shown in Figure 2. The 
color distribution of objects in M82 South runs from $r'-i' \geq -0.3$ to 0.5, while the LF 
indicates that the brightest stars in M82 South have $i' \sim 23$.

	The (M$_{i'}, (r'-i')_0$) CMD of stars within $\pm 20$ arcsec of the 
M82 South ridgeline is shown in Figure 3. A distance modulus of 27.95 (Sakai \& Madore 
1999) has been adopted, with a foreground reddening A$_B = 0.10$ (Burstein \& 
Heiles 1984). There is dust in the M82 outflow (e.g. Heckman, Armus, \& Miley 1990), and so 
the foreground reddening is a lower limit to the actual reddening. This being said, 
M82 South is much further from the disk plane than the area where dust has been detected. 
Furthermore, the mean colors and color distributions of stars in the eastern and western 
portions of M82 South are not different, indicating that if dust is present then it 
is very uniformly distributed.

	Also shown in Figure 3 are evolutionary tracks from Girardi et al. (2004) 
with Z = 0.008 and Z = 0.019 for ages log(t$_{yr}$) = 7.5 and 8.0. 
While not shown in Figure 3, isochrones with Z = 0.0001 place the red supergiant locus at 
$r'-i'$ colors that are $\sim 0.2 - 0.3$ smaller than those with Z = 0.008. With the 
caveat that up to one third of the stars probably do not belong to M82 South, 
then the red envelope of stars in Figure 3 argues for Z $\geq 0.008$, which is consistent 
with the metallicity of young stars in the disk of M82 (e.g. Mayya et al. 2006). 
The stars in M82 South thus formed from chemically enriched gas. 

	The ages of the youngest stars can also be estimated from the comparisons in 
Figure 3. The large number of stars with M$_{i'}$ between --4 and --5 
and $r'-i'$ between --0.1 and 0.3 is consistent with log(t$_{yr}$) = 7.5 -- 8.0; 
thus, log(t$_{yr}$) = 7.75 is adopted for the youngest stars.
An interesting check of this age comes from a comparison with Ho IX, which has an SED 
at visible wavelengths that is very similar to that of M82 South (Sun et al. 2005). 
Makarova et al. (2002) find that the majority of stars in Ho IX have ages log(t$_{yr}) 
= 7.8$, in excellent agreement with what is found in M82 South. 

\section{DISCUSSION}

	Deep images obtained with the CFHT MegaCam have been used to resolve individual 
stars in M82 South. The youngest stars have ages log($t_{yr}) = 7.75$ and Z $\geq 0.008$. 
Thus, despite having a projected distance of $\sim 6$ kpc off of the M82 disk plane, 
M82 South recently formed stars from chemically enriched material.

	Lacking kinematic information, the physical separation between M82 South and 
the main body of M82 is a matter of speculation. M82 South is not in the disk plane of M82, 
and it seems likely that M82 South is associated with the extraplanar regions of M82, as 
opposed to being an outlying structure on the far or near side of the M81 group, given its 
close projected proximity to M82. The general morphology of M82 South in Figure 4 
of Sun et al. (2005) is reminiscent of the filamentary features that originate 
from M81, albeit on a much smaller scale. Thus, M82 South 
physically resembles features seen in the outer regions of other galaxies. 

	Adopting the mass estimates in \S 3 then 
it is unlikely that M82 South will be a long-lived feature. 
If the mass of M82 is 10$^{10}$ M$_{\odot}$ (Sofue et 
al. 1992), then the tidal radius of M82 South is a few tenths of a kpc if the projected 
separation of 6 kpc corresponds to the actual separation. To escape tidal pruning, M82 
South would have to be 50 -- 100 kpc away from M82, on the side of M82 that is furthest 
from M81. If it is within 10 -- 20 kpc of M82 then M82 South will survive for only a few 
orbital crossing times about M82, or $\sim 10^8 - 10^9$ years, depending on the actual 
distance from M82 and the nature of the orbit about the larger galaxy.

	Could M82 South be the remnant of a pre-existing dwarf galaxy that was disrupted 
by M82? This is unlikely, as the metallicity of stars in M82 South suggest that 
such a satellite would have had an LMC-like mass, which is roughly 25 -- 50\% that of 
M82. The disruption of such a massive galaxy would leave significant tidal debris 
trails, which are not seen. 

	There are conspicuous signatures of galaxy-galaxy interactions throughout the 
central regions of the M81 Group. Gas bridges link galaxies (e.g. Yun et al. 1994; Boyce 
et al. 2001), and there are objects that may be tidal dwarfs (Karachentsev 
et al. 2002; Makarova et al. 2002). Could M82 South be such a 
tidal fragment? The youngest stars in M82 South have ages near 
50 Myr, and there was contemporaneous star formation in other 
candidate tidal features, such as Ho IX, the Garland, and the Arp-Loop (e.g. Makarova 
et al. 2002; Sakai \& Madore 2001). However, tidal dwarfs are predicted to be gas-rich 
(e.g. Barnes \& Hernquist 1992), and the HI map of Yun et al. (1994) does not show an 
HI concentration near M82 South, which is in marked contrast to Ho IX and the Arp-Loop.
This being said, the presence of diffuse H$\alpha$ emission (Figure 2 
of Hoopes et al. 2005) suggests that M82 South may not be completely devoid of gas. 

	The outflow from M82 has ejected at least $3 \times 10^8$ M$_{\odot}$ of gas out 
of the disk plane  (Walter et al. 2002), and M82 South may have condensed out of 
some of this material. The `Cap' (Devine \& Bally 1999) is 
located some 11 kpc to the north of the M82 disk plane, and is 
thought to have formed in a shock that occured when the outflow 
encountered clouds surrounding M82 (e.g. Lehnert et al. 1999; Strickland et al. 2004). 
The material in the Cap is metal-enriched (e.g. Hoopes et al. 2005; Strickland \& Heckman 
2007; Tsuru et al. 2007), and there are knots that may indicate the onset 
of star formation, although stars have yet to be resolved. If M82 South is a related 
structure then it must be in a later stage of evolution. M82 South is a much weaker 
source of x-ray (e.g. Tsuru et al. 2007) and H$\alpha$ (Hoopes et al. 2005) emission than 
the Cap, but is a source of much stronger UV emission, which is concentrated in at least 
three knots (Hoopes et al. 2005). While the absence of detectable x-ray emission 
indicates that the material in M82 South is not at present experiencing shock excitation, 
such activity in the past might have triggered star formation. The drop in stellar density 
that defines the western edge of M82 South might then mark the spatial extent 
of star-forming material.

	If M82 South is a genuinely young object 
then it will not have an underlying population of old stars. Deep high angular resolution 
imaging will provide constraints on objects such as stars evolving on the 
red giant branch; if such stars are present then they would indicate an age of at least 
$\sim 1$ Gyr for M82 South. The detection of such stars would suggest that M82 South 
is probably more distant from M82 than 6 kpc, given the disruption time estimated 
above. 

	We conclude by noting that if structures like M82 South 
are common in interacting galaxies then they may have a significant impact on the 
extraplanar stellar content of these objects. If M82 South is dispersing then its young 
metal-rich stars may eventually migrate into the extraplanar regions of M82 or the 
intragalactic medium in the M81 group. If similar structures have formed previously and 
dispersed near M82 then deep imaging of the outer regions of M82 should reveal stars with a 
wide range of ages and metallicities. The CMDs of the north east control field in Figure 2 
and in the extraplanar regions of other galaxies (e.g. Mouchine 2006) suggest that such a 
dispersion in stellar content is present.

\acknowledgements{Sincere thanks are extended to Brenda Matthews for discussions regarding 
the role of shocks in star formation. It is also a pleasure to thank the anonymous referee 
for providing comments that greatly improved the manuscript.}

\clearpage

\begin{figure}
\figurenum{1}
\epsscale{0.85}
\plotone{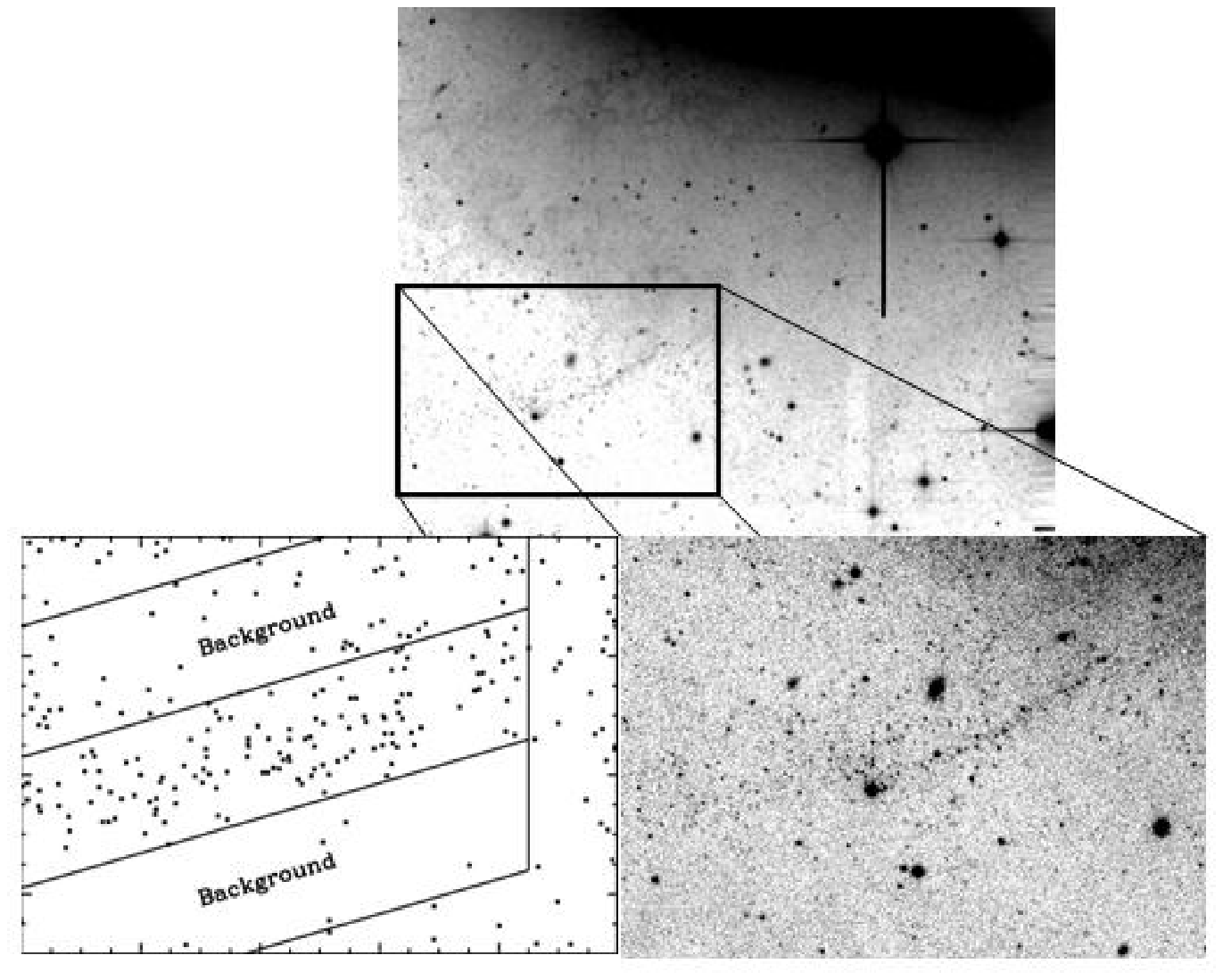}
\caption
{The top portion of this figure shows a $370 \times 370$ arcsec$^2$ section of 
the final processed $i'$ MegaCam image, with north at the top and east to the left. The 
gradient in surface brightness due to stars in M82 is clearly seen. 
The $180 \times 150$ arcsec$^2$ area around M82 South 
is shown in the lower right hand corner, while the spatial distribution of stars with 
$i'$ between 22.5 and 24.5 in this same area is shown in the left hand inset. 
The regions used in the analysis that sample the main concentration of stars in M82 South 
and the background/control fields are indicated.}
\end{figure}

\clearpage

\begin{figure}
\figurenum{2}
\epsscale{0.85}
\plotone{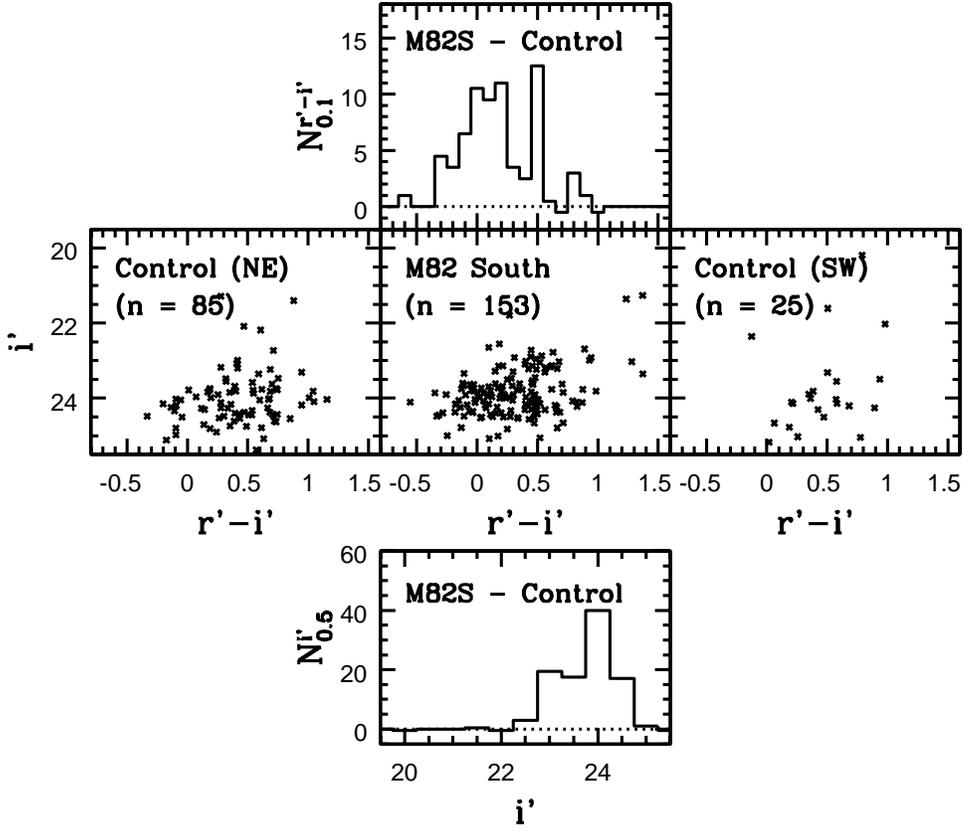}
\caption
{The $(i', r'-i')$ CMD of objects in a $\pm 20$ arcsec region 
centered on the ridgeline of M82 South and in control fields 
located $\pm 40$ arcsec on either side of M82 South 
are shown in the middle row. The number of objects in each CMD is listed, 
and there is a clear excess of sources associated with M82 South. 
The result of subtracting the mean $r'-i'$ color distribution of 
objects with $i'$ between 23.5 and 24.5 (M$_{i'} \sim -4.5$ to --3.5) in the control fields 
from the color distribution of sources in M82 South is shown in the top row, where 
N$^{r'-i'}_{0.1}$ is the net number of stars in this brightness interval per 0.1 
magnitude $r'-i'$ interval. Note that the color distribution runs from $r'-i' = --0.3$ 
to 0.5, with a peak near $r'-i' \sim 0$. The LF that results from subtracting the mean LF 
of the control fields from the LF of M82 South is shown in the bottom panel, where 
N$^{i'}_{0.5}$ is the net number of stars per 0.5 magnitude $i'$ interval. Note that the 
majority of stars in M82 South have $i' \geq 23.0$, which corresponds roughly to 
M$_{i'} \geq -6$.}
\end{figure}

\clearpage

\begin{figure}
\figurenum{3}
\epsscale{0.95}
\plotone{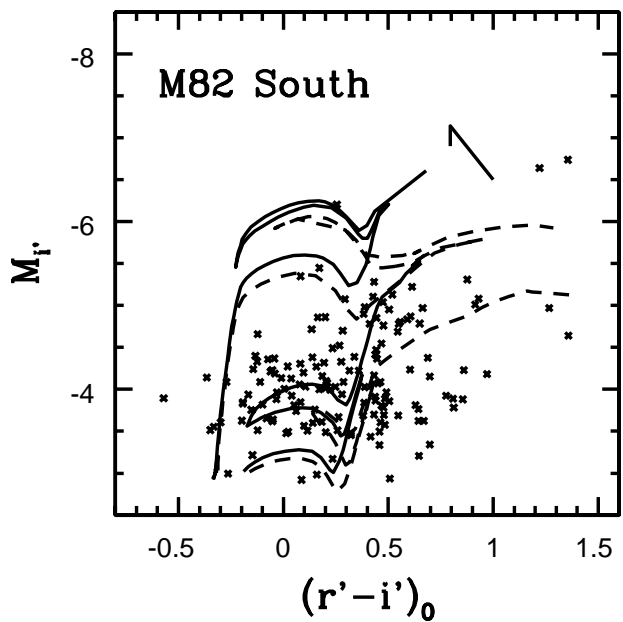}
\caption
{The $(M_{i'}, (r'-i')_0)$ CMD of the $\pm 20$ arcsec region centered on M82 South. 
A reddening vector with a length that is appropriate for A$_V = 1$ is shown.
The solid lines are Z = 0.008 isochrones with ages log(t$_{yr}$) = 7.5 and 8.0 
from Girardi et al. (2004), while the dashed lines are isochrones with the same ages, but 
Z = 0.019. Note that the brightnesses of stars with $(r'-i')_0$ between 
--0.3 and 0.5 favours an age between log(t$_{yr}$) = 7.5 and 8.0, while the red envelope of 
stars suggests that $Z \geq 0.008$.}
\end{figure}

\end{document}